\documentclass[aps,pre,reprint, superscriptaddress]{revtex4-1}

\usepackage{graphicx}% Include figure files
\usepackage{dcolumn}% Align table columns on decimal point
\usepackage{bm}% bold math

\begin{document}

\title{What Can We Learn from Noise?\\ -- Mesoscopic Nonequilibrium Statistical Physics -- }
\author{Kensuke Kobayashi}
\affiliation{Graduate School of Science, Osaka University, 
1-1 Machikaneyama, Toyonaka, Osaka 560-0043, Japan.}

\begin{abstract}
Mesoscopic systems -- small electric circuits working in quantum regime -- offer us a unique experimental stage to explorer quantum transport in a tunable and precise way. The purpose of this Review is to show how they can contribute to statistical physics. We introduce the significance of fluctuation, or equivalently noise, as noise measurement enables us to address the fundamental aspects of a physical system. The significance of the fluctuation theorem (FT)  in statistical physics is noted. We explain what information can be deduced from the current noise measurement in mesoscopic systems.  As an important application of the noise measurement to statistical physics,  we describe our experimental work on the current and current noise in an electron interferometer, which is the first experimental test of FT in quantum regime. Our attempt will shed new light in the  research field of mesoscopic quantum statistical physics.
\end{abstract}

\maketitle 
%==========================================================================

\section{Introduction}
Mesoscopic systems are very small, typically micrometer- or nanometer- sized, electric circuits that are made of metals or semiconductors by using microfabrication technique~\cite{DattaETMS,ImryIMP,IhnEQTMSS}. The biggest advantage of studying them lies in the fact that we can conduct precise experiments in a scale where quantum physics plays a key role. Actually, many phenomena based on charge, spin, coherence and many-body effects of electrons have been demonstrated in various mesoscopic systems. For example, the wave nature of electrons can be controlled in electron interferometers such as  Aharonov-Bohm (AB) rings, while the particle nature of electrons can be tuned in artificial atoms or quantum dots. Since 1980's, many researchers have been intensively working to understand and control a variety of quantum effects, which has greatly contributed to deepening our understanding of quantum physics and solid state physics. 

Most experimental studies on mesoscopic systems have been performed by using the conductance measurement. This is a matter of course, not only because the conductance is the most basic physical value to characterize the circuit, but also because the conductance in a mesoscopic system directly reflects the transmission probability of the system according to the Landauer formula, as we explain later. The conductance (or current) is the time-averaged value; the measured value stands for the average number of electrons that pass through a given system for a finite time. On the other hand, the fluctuation of the current (``current noise''), namely the fluctuation of the number of electrons passing through the system, conveys us essentially different information. Theoretically, the current noise has been regarded as an important topic in mesoscopic physics and a lot of theoretical work has been done since early 1990's~\cite{BlanterPR2000,BeenakkerPT2003}. In spite of great theoretical interest, there have been a comparatively small number of experimental studies available in the literatures. One of the reasons is that the noise measurement is technically difficult when compared to the conventional conductance measurement. Nevertheless, as we describe below, the demonstration of the fractional charge in the fractional quantum Hall regime probed  in the noise measurement~\cite{de-PicciottoNature1997,SaminadayarPRL1997} is widely appreciated.  As exemplified by this work, the noise measurement is unique as it gives information that would be impossible in other measurement techniques. 

These few years, we are interested in noise in mesoscopic systems and have intensively performed the noise measurement in various systems, for example, quantum point contact~\cite{HashisakaPRB2008,NakamuraPRB2009,NishiharaAPL2012,KohdaNatComm2012,MuroPRB2016}, quantum wire~\cite{ChidaPRB2013}, electron interferometer~\cite{HashisakaPE2010,NakamuraPRL2010,NakamuraPRB2011}, quantum dot~\cite{YamauchiPRL2011,FerrierNatPhys2016}, magnetic tunnel junction~\cite{SekiguchiAPL2010,ArakawaAPL2011,ArakawaPRB2012,%
TanakaAPEX2012,TanakaAPEX2014}, spin valve~\cite{ArakawaPRL2015}, and graphene~\cite{MatsuoNatComm2015,TakeshitaAPL2016}.  In this Review, we discuss that the noise measurement can reveal the fundamental  aspects of physical systems. Especially, we emphasize its impact on statistical physics.

This Review is organized as follows. First, in Sec.~\ref{IntroNoise}, we introduce what is noise (fluctuation) and then explain ``fluctuation theorem'' (FT)~\cite{EvansPRL1993}, which plays an important role in statistical physics since its discovery in 1990's. Second, in Sec.~\ref{IntroMesoNoise}, we describe the noise in mesoscopic systems by giving several examples. Third, in Sec.~\ref{FTtest}, after discussing FT in mesoscopic transport, we explain our work on the experimental test of FT in mesoscopic systems. By applying FT to mesoscopic transport, we have established non-trivial relations between nonequilibrium fluctuation and nonlinear response of the system, which is beyond the well-known fluctuation-dissipation relations. Finally, in Sec.~\ref{SecConcPersp}, the concluding remarks and future perspectives are presented.

\section{What Fluctuation Indicates}
\label{IntroNoise}

\subsection{Fluctuation-Dissipation (FD) Relations}
Very generally, to describe a physical system, we first investigate how it responds to an external field. For example, we measure its electric resistance (or conductance), its magnetization, its dielectric constant, and so on in order to see the response of the system to the electric field and the magnetic field. We are familiar with Ohm's law, which tells us that the electric current through a conductor is proportional to the bias voltage applied to it. 

\begin{figure}[tbp]
\center\includegraphics[width=.99\linewidth]{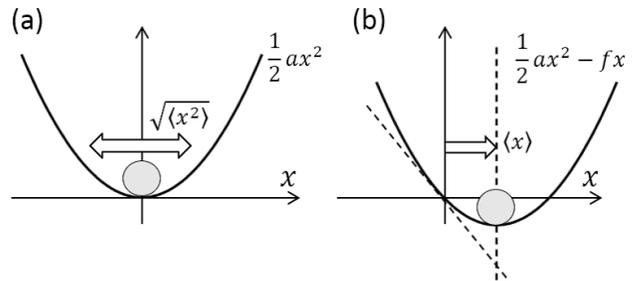}
\caption{(a) A particle that is trapped in a one-dimensional harmonic potential. Its position is thermally fluctuating. (b) The particle position is shifted when the external field $f$ is applied.} 
\label{KirkwoodFig}
\end{figure}

Now, we discuss that the response of the physical system and the fluctuation that resides in it are closely connected. We show the simplest example~\cite{KirkwoodJCP1946}. Consider a classical particle that is trapped in a one-dimensional harmonic potential as shown in Fig.~\ref{KirkwoodFig} (a). The Hamiltonian is given as follows,
\begin{equation}
{\cal H} = \frac{1}{2}a x^2,
\end{equation}
where $x$ is coordinate and $a>0$.
The particle should be at the origin ($x=0$). When the force $f$ is applied to it as shown in Fig.~\ref{KirkwoodFig} (b), the Hamiltonian changes to 
\begin{equation}
{\cal H} = \frac{1}{2}a x^2 - fx. 
\end{equation}
The position of the particle is shifted from the origin to be 
\begin{equation}
\langle x \rangle =\frac{f}{a}.
\end{equation}
Here, we define $\langle \cdots \rangle$ to represent the expected value of $\cdots$.  Now even if there is no external field $f$, the particle is thermally fluctuating around the origin as shown in Fig.~\ref{KirkwoodFig} (a). The amplitude of the fluctuation $\langle x^2 \rangle$ at the environment temperature $T$ can be calculated as
\begin{equation}
\langle x^2\rangle = 
\frac{\int^{\infty}_{-\infty}dx x^2 e^{-\frac{1}{2}ax^2/k_BT}}{\int^{\infty}_{-\infty}dx  e^{-\frac{1}{2}ax^2/k_BT}} = \frac{k_BT}{a}=\langle x \rangle\frac{k_BT}{f}.
\label{xsquare}
\end{equation}
When we define the susceptibility or response coefficient $\chi$ of the particle for the applied force $f$ as $ \langle x\rangle \equiv \chi f$, 
\begin{equation}
\chi = \frac{\langle x^2\rangle}{k_BT} \label{Kirkwood}
\end{equation}	
holds because of  Eq.~(\ref{xsquare}). Eq.~(\ref{Kirkwood}) is called Kirkwood relation~\cite{KirkwoodJCP1946}. 
It suggests that the thermal fluctuation of the particle 
$\langle x^2\rangle$ characterizes how it responds to the field. 
This interesting relation is not just a coincidence. Actually, such a relation indicating ``fluctuation tells response'' is general and is widely seen in many physical systems. This is called the fluctuation-dissipation (FD) relations~\cite{KuboRPP1966}. 

The FD relations first appeared in Einstein's Brownian motion theory in 1905~\cite{EinsteinAP1905}. He considered the thermal fluctuation of a particle in liquid, which is subject to continuous collisions of the surrounding molecules. He derived the linear relation between the diffusion constant and the mobility of the particle (Einstein's relation). Based on his theory, Perrin precisely measured the Avogadro constant in 1909 and showed that molecules are real entity~\cite{PerrinBMMR1910}, for which he was awarded the Nobel prize in physics in 1926. Einstein's theory and Perrin's experiment beautifully prove that we can learn profound essence of nature by carefully observing the fluctuation. 

The FD relations are widely seen in various systems~\cite{CallenPR1951,GreenJCP1954,KuboJPSJ1957}. We will discuss another typical example, the thermal noise of a resistor~\cite{JohnsonPR1928,NyquistPR1928} below. In 1950's, such observations were generalized into the linear-response theory, or so-called Kubo formula~\cite{KuboJPSJ1957}. Nowadays, this beautiful formalism is established as the most standard tool for investigating the properties of physical systems. 

\subsection{Fluctuation Theorem (FT)}

Although the formalism of the linear-response theory is powerful, we have to note that it is only applicable when the system is around the equilibrium or at most within the linear-response regime. We usually encounter various phenomena, which are essentially nonequilibrium and nonlinear, such as light-matter interaction, chemical reaction, electronic circuit with active components like transistors, life activity, and so on. Tremendous efforts, therefore, have been paid in order to overcome the limit of linear-response theory and to understand nonequilibrium systems. 

\begin{figure}[tbp]
\center\includegraphics[width=.99\linewidth]{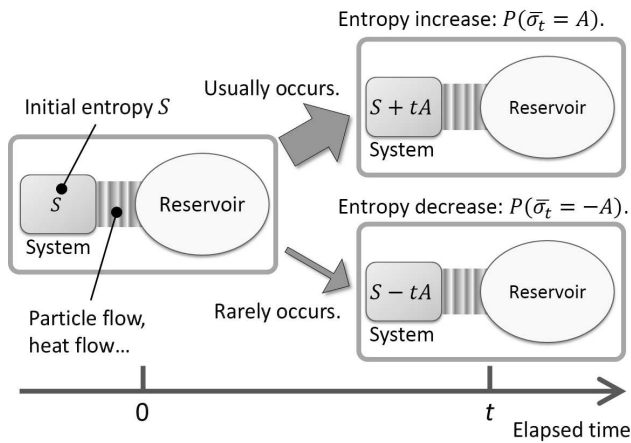}
\caption{Schematic view of FT. Consider a small system coupled to the reservoir. There exists flow such as particle flow or heat flow between the two, and thus the entropy in each system varies with time. We define the initial entropy of the small system to be $S$ at time 0 and consider the probability   that it has the entropy of $S+tA$ at a certain time $t>0$ and the probability with the entropy of $S-tA$ at time $t$ (we take $A>0$). These two probabilities are exactly linked according to Eq.~(\ref{FTeqn}).  } 
\label{FTpicture}
\end{figure}

One of the most successful outcomes along this direction is the fluctuation theorem (FT) proposed by Evans, Cohen, and Morriss in 1993~\cite{EvansPRL1993}. As shown in Fig.~\ref{FTpicture}, consider a small system that is coupled to the reservoir. We assume that there exists flow (particle flow, heat flow, and so on) between the two, and the entropy in each system varies with time. While the second law of thermodynamics tells that the total entropy should increase, that of the small system may fluctuate. Based on microscopic reversibility (``micro reversibility'' or detailed balance), FT exactly links the probabilities of the production and consumption of the entropy in the system. 
We define the entropy production rate of the small system $\sigma(t)$ and its time-averaged value for a finite time $t$ as $\displaystyle \overline{\sigma}_t \equiv \frac{1}{t}\int_0^t ds \sigma (s)$. Then the probability $P(\overline{\sigma}_t)$, which defines the probability of the entropy production rate $\overline{\sigma}_t$, exactly satisfies the relation~\cite{EvansAP2002},
\begin{equation}
\frac{P(\overline{\sigma}_t=A)}{P(\overline{\sigma}_t=-A)}=\exp (t A).
\label{FTeqn}
\end{equation}
According to this expression, the entropy should increase for a long time as the right-hand side of Eq.~(\ref{FTeqn}) becomes large for $t\rightarrow \infty$ when $A>0$. For a short time, in contrast, the entropy can either increase or decrease. Thus, this representation may be viewed as a quantitative description of the second law of thermodynamics. Furthermore, Eq.~(\ref{FTeqn}) is unique as it strictly connects between the probability of very rare events with $P(\overline{\sigma}_t=-A)\ll 1$ and the most ordinary events with $P(\overline{\sigma}_t=A)\sim 1$. FT was also proven to reproduce the linear-response theory~\cite{GallavottiPRL1996}, and the Onsager-Casimir relations~\cite{SaitoPRB2008,UtsumiPRB2009}.

So far, FT has been experimentally addressed in several ways~\cite{BustamantePT2005}, for example, in the investigation of the movement of a single micro particle in a fluid~\cite{WangPRL2002}, the noise in a bulk resistor~\cite{GarnierPRE2005}, and the electron counting experiment in a quantum dot~\cite{UtsumiPRB2010,KungPRX2012}. While these are the experiments in the classical regime, there had been no experimental attempt to investigate whether or not FT is applicable to situations in the quantum regime. The purpose of this Review is to explain how to fill this gap. In 2010, we tackled this issue by using our precise current noise measurement system~\cite{NakamuraPRL2010,NakamuraPRB2011}. In the next Section, we discuss the current noise, and then we will come back to this topic in Sec.~\ref{FTtest}.

%==========================================================================

\section{Current Fluctuation in Mesoscopic Systems}
\label{IntroMesoNoise}

\subsection{Current Noise}
\begin{figure}[bp]
\center\includegraphics[width=.99\linewidth]{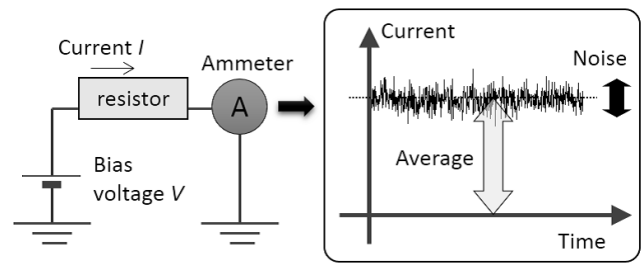} 
\caption{Measurement scheme of the current and the current noise. A constant bias voltage $V$ is applied to a resistor and the electric current $I$ that flows through it is precisely monitored as a function of time by an ammeter. The current usually has a finite noise $\langle (\delta I)^2 \rangle$ around its averaged value $\langle I \rangle$.  }
\label{currentmeasurement}
\end{figure}

First of all, we discuss the definition and some basic properties of the noises. As shown in Fig.~\ref{currentmeasurement}, consider the situation that we connect a resistor to a voltage source and apply a bias voltage $V$. By using an ammeter, we precisely monitor the electric current $I$ that flows through the resistor. Of course, the current has a time-averaged value $\langle I \rangle$ but it also shows a finite fluctuation, or equivalently noise, and usually $\delta I \equiv I -\langle I \rangle \neq 0$. The noise spectral density $S$ is defined through the mean square of $\delta I$ for the band width $\Delta f$, 
\begin{equation}
S   =  \frac{\langle (\delta I)^2 \rangle}{\Delta f}.
\end{equation}
The unit of this value $S$ is $[\mbox{A}^2/\mbox{Hz}]$ as is easily seen from this definition. To examine  physical units is instructive. As the current has the unit of $[\mbox{A}]=[\mbox{C}/\mbox{s}]$, it corresponds to the number of electrons that pass through the system for a finite time. On the other hand, as $[\mbox{A}^2/\mbox{Hz}] = [\mbox{C}^2/\mbox{s}]$, the current noise amplitude stands for the variance of the number of electrons that pass through the system for the corresponding time. 

When $V=0$, the system is in the equilibrium and $\langle I \rangle=0$. This, however, does not necessarily mean that there is no noise. Actually, at finite temperature, there arises finite noise, so-called ``thermal noise'' or  Johnson-Nyquist noise, which is expressed as
\begin{equation}
S_{\mbox{\small th}} = 4 k_B T G,
\label{JN}
\end{equation}
where $G$, $T$, and $k_B$ are the conductance of the resistor, its temperature, and the  Boltzmann constant, respectively~\cite{JohnsonPR1928,NyquistPR1928}. $G$ represents a kind of response of the resistor, as this value quantifies the current that flows through the resistor when the bias voltage $V$ is applied  ($I =GV$). Therefore, Eq.~(\ref{JN}) is a typical example of the FD relation telling us that the fluctuation tells the response. This analogy can be seen when we compare Eq.~(\ref{Kirkwood}) and Eq.~(\ref{JN}). 

In 1928, Nyquist elegantly deduced Eq.~(\ref{JN}) based on the second law of thermodynamics~\cite{NyquistPR1928} to explain the experimental observation of Johnson~\cite{JohnsonPR1928}. He also pointed out the relation between this formula and the black-body radiation. These facts clearly  indicate that fundamental physics underlies noise.

For the case of $V \neq 0$, namely in the nonequilibrium case, the situation totally changes. For simplicity, consider a device that is a thin potential barrier sandwiched by two metallic leads, such as realized in metal/insulator/metal junctions or $pn$ junctions. The electrons can be either transmitted or reflected at the barrier. Due to such stochastic processes, the current passing through the device fluctuates inevitably. This noise, which is called ``shot'' noise, directly reflects the granularity of electrons that carry current. In 1918, Schottky gave the expression of the shot noise as~\cite{SchottkyAP1918}
\begin{equation}
S_{\mbox{\small shot}}=2e\langle I \rangle.
\label{Schottky}
\end{equation}
The reason for this formula can be explained in the following way. The current is expressed as $\langle I \rangle=e\langle Q \rangle/\tau$, where $\langle Q \rangle/\tau$ is the number of electrons that pass through the device ($Q$) averaged over for a finite time $\tau$. We assume that there is no correlation between each tunneling event passing through the barrier, namely that the tunneling process obeys Poisson distribution. Then according to the basic property of this distribution, the variance of the number of the electrons that pass through the barrier equals to the average number of them; $\langle (\delta Q)^2 \rangle \equiv \langle (Q - \langle Q \rangle)^2 \rangle=\langle Q \rangle$. Therefore, the current noise should satisfy $\langle (\delta I)^2
\rangle= e^2 \langle (\delta Q)^2 \rangle /\tau = e^2 \langle Q \rangle /\tau = e\langle I
\rangle$, which is nothing but Eq.~(\ref{Schottky}) (note that the factor 2 is customarily put in Eq.~(\ref{Schottky}) without serious reason).

The functional forms of the equilibrium noise (Eq.~(\ref{JN})) and the nonequilibrium noise (Eq.~(\ref{Schottky})) are different between each other. This vividly reflects that the nonequilibrium properties of the system are different  from its equilibrium ones in a nontrivial way. Especially, we emphasize that the elementary charge $e$ does not appear in the equilibrium but does in the nonequilibrium as seen in  Eq.~(\ref{Schottky}). As a historical remark, the nonequilibrium noise~\cite{SchottkyAP1918}, namely the shot noise was discovered 10 years earlier than the equilibrium  noise~\cite{JohnsonPR1928,NyquistPR1928}, which might signal that our daily life is essentially nonequilibrium. 

In the above discussion, we have implicitly assumed the noise in the low frequency limit, where both $S_{\mbox{\small th}}$ and $S_{\mbox{\small shot}}$ are frequency-independent and thus called ``white'' noises. In this article, we focus on these two kinds of noise. In the applied physics research fields, other noises such as the $1/f$ noise and the telegraph noise are also important as these affect the properties of electronic devices~\cite{DuttaRMP1981}. These noises, however, are highly dependent on the frequency range and are device-dependent and therefore are difficult to treat quantitatively.

\subsection{Landauer-B{\"u}ttiker Framework}
Transport in mesoscopic systems has been successfully treated in the framework of Landauer-B{\"u}ttiker formalism~\cite{DattaETMS,ImryIMP,IhnEQTMSS}. The current noise is also possible to be dealt within this framework~\cite{BlanterPR2000}. While we do not explain the theoretical derivation, we briefly summarize the formula to show the shot noise in mesoscopic systems (Eq.~(\ref{ShotTheory})) here. Although it requires several assumptions, this formula has been successfully applied in many experiments on the current noise in mesoscopic systems. We show several examples just later in Sec.~\ref{subsec:exampleofnoise}.

First, the well-known Landauer formula for the conductance is described~\cite{DattaETMS,ImryIMP,IhnEQTMSS,BlanterPR2000}. Consider a mesoscopic conductor that consists of $n$ conducting channels, where a transmission probability of the $n$-th channel is defined $T_n$. 
The conductance $G$ of the conductor is given by the following expression, 
\begin{equation}
 G = \frac{2e^2}{h}\sum_n T_n.
\label{Landauer}
\end{equation}
This is called Landauer formula. ${2e^2}/{h}$ is the conductance quantum $\sim (12.9~{\rm k\Omega})^{-1}$, twice the reciprocal of the von Klitzing constant.

\begin{figure}[tbp]
\center\includegraphics[width=.7\linewidth]{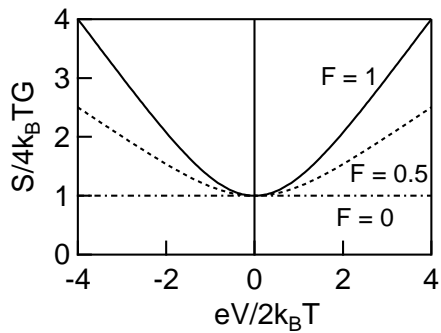}
\caption{Dimensionless value of $S/4k_BTG = 1+F(X\coth(X)-1)$ (Eq.~(\ref{ShotTheory})) is plotted as a function of $X \equiv eV/2k_BT$ for $F=0, 0.5$, and $1$.}
\label{shotmodel}
\end{figure}

When the scale of the energy dependence of $T_n$ is much larger than both the temperature $T$ and the bias-voltage $V$, the noise spectral density $S$ in the low frequency limit can be expressed as~\cite{BlanterPR2000,ButtikerPRL1990,ButtikerPRB1992}
\begin{equation}
S = 4 k_BTG + 2e\langle I \rangle F
 \left[  \coth\left(\frac{eV}{2k_BT}\right) - \frac{2k_BT}{eV} \right].
\label{ShotTheory}
\end{equation}
This is  independent of frequency.  $F$ is called Fano factor which is defined as 
\begin{equation}
F \equiv \frac{\sum_n T_n(1-T_n)}{\sum_n T_n}.
\label{FanoTheory}
\end{equation}
As shown in Fig.~\ref{shotmodel}, Eq.~(\ref{ShotTheory}) normalized by $4k_BTG$ is the function of $eV/2k_BT$. As is clear from the figure, $S = S_{\mbox{\small th}}=4 k_BTG$ holds for $\vert eV\vert \ll 2k_BT$ and $S = S_{\mbox{\small shot}}\times F$ for another limit of $\vert eV\vert \gg 2k_BT$. In this way, the thermal noise and the shot noise are smoothly connected as a function of $eV/2k_BT$ in the Landauer-B{\"u}ttiker framework. 

The Fano factor $F$ satisfies $F = S/2e\langle I \rangle$ for $\vert eV\vert \gg 2k_BT$. It tells whether or not the electron transport is Poissonian. $F>1$ and $F<1$ correspond to the super-Poissonian and the sub-Poissonian cases, respectively. Such information is useful because it signals how correlated electrons are. For $F=1$, we can safely say that each tunneling event is independent of each other. On the other hand,  $F>1$ and $F<1$ correspond to electron bunching or anti-bunching, respectively. 

\subsection{Examples of Current Noise in Mesoscopic Transport}
\label{subsec:exampleofnoise}
\begin{figure}[tbp]
\center\includegraphics[width=.95\linewidth]{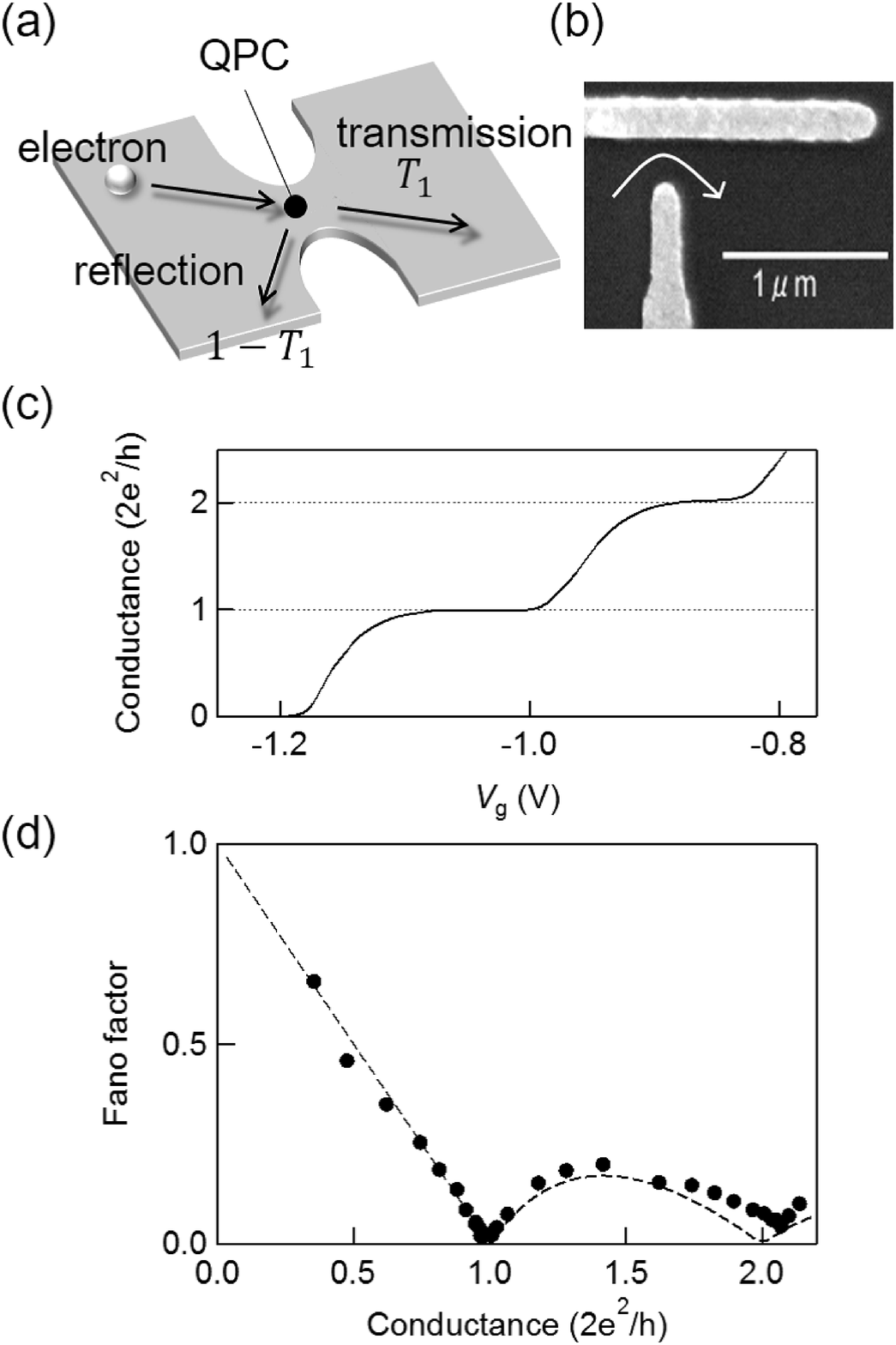}
\caption{(a) Schematic picture of the QPC. The simplest case with a single conducting channel is shown. The transmission and reflection probabilities are tuned to be $T_1$ and $1-T_1$, respectively. (b)  Scanning electron microscope image of the QPC fabricated on the GaAs/AlGaAs two-dimensional electron gas (2DEG) system~\cite{NakamuraPRB2009}. (c) Conductance of the QPC as a function of the gate voltage $V_g$. (d) Fano factor as a function of the conductance $G$.  }
\label{QPC_Fano}
\end{figure}

Here we show several examples of the current noise study. First, we show the experimental result of the quantum point contact (QPC). We show a schematic picture of the QPC in Fig.~\ref{QPC_Fano}(a). This is a  small conductor as narrow as the Fermi wave length of electrons. As explained below, we can electrostatically tune the width of this narrow path and thus tune the number of conducting channels and the transmission probability of each channel. Figure~\ref{QPC_Fano}(a) shows the simplest case; only a single conducting channel is formed where the transmission and reflection probabilities are tuned to be $T_1$ and $1-T_1$, respectively. Therefore the conductance is given by $G = \frac{2e^2}{h}T_1$ and the Fano factor is $F = 1-T_1$ according to Eq.~(\ref{FanoTheory}). The shot noise theory in the QPC was intensively studied  by several theorists at the early stage of mesoscopic physics~\cite{LesovikJETP1989,YurkePRB1990,ButtikerPRL1990,MartinPRB1992,ButtikerPRB1992}.

Figure~\ref{QPC_Fano}(b) shows  a scanning electron microscope (SEM) image of the QPC fabricated on the GaAs/AlGaAs two-dimensional electron gas (2DEG) system~\cite{NakamuraPRB2009}.
The QPC is defined by the two metallic gates and by changing the voltage applied to these gates ($V_g$) the width of the QPC is electrostatically controlled. As the Fermi wavelength is $\sim 50$~nm for an ordinary 2DEG, to control the number of the conducting channels ($n$) and their transmission probability ($T_n$) is possible in the QPC. As a result, by sweeping the gate voltage, the QPC shows the quantized conductance at $G = 2e^2/h$ and $G = 4e^2/h$ as shown in Fig.~\ref{QPC_Fano}(c). This is a beautiful manifestation of the Landauer formula (Eq.~(\ref{Landauer})).

Figure~\ref{QPC_Fano}(d) represents the measured Fano factor as a function of the conductance $G$. The black circles show the result of the analysis of the experimental result based on Eq.~(\ref{ShotTheory}) (details of the noise experiments will be described later). The dashed curve is the theoretical Fano factor given by Eq.~(\ref{FanoTheory}). They satisfactorily agree well each other. Very importantly, at the quantized conductance of $2e^2/h$ and $4e^2/h$, the expected Fano factors are zero and the observed ones are close to zero as theoretically predicted~\cite{BlanterPR2000,LesovikJETP1989,YurkePRB1990,ButtikerPRL1990,MartinPRB1992,ButtikerPRB1992}. This indicates that when there is a perfectly transmitting channel, no electrons can be reflected since there are no empty states for them to be reflected back to, leading to the absence of fluctuations. Thus this observation is the direct consequence of Pauli exclusion principle. At the same time, the above observation (perfect anti-bunching) also guarantees that the electron transport occurs in a quantum coherent way. This simple example clearly teaches us the advantage of the noise measurement in mesoscopic systems.  

Recently, we performed a precise Fano factor measurement of the MgO-based magnetic tunnel junction (MTJ) devices~\cite{SekiguchiAPL2010,ArakawaAPL2011,ArakawaPRB2012,TanakaAPEX2012}. By determining the Fano factor within 1\% precision, we established that when the tunneling barrier is very thin, the Fano factor is less than the unity, namely, the Poissonian value. The suppressed Fano factor indicates the relevance of the coherent transport even in the tunnel junction. Actually, our result quantitatively supports a ``coherent tunneling model'' of the MgO-based MTJs~\cite{ArakawaAPL2011,ArakawaPRB2012} as reproduced by the theory~\cite{LiuPRB2012}. 

The most well-known current noise study in mesoscopic physics is the experiments that proved the relevance of the fractional charge. If the Schottky's argument can be applied to the quasi-particle excitation with the effective charge $e^* \neq e$, the shot noise would be expressed as $S=2e^*\langle I \rangle$ after Eq.~(\ref{Schottky}). Actually, relying on this fact, the researchers successfully demonstrated that the carriers in the fractional quantum Hall regime have the fractional charges of $e^*=e/3$~\cite{de-PicciottoNature1997,SaminadayarPRL1997}. From the conventional current (or conductance) measurements, it is impossible to discuss the nature of the carriers which convey electric current. The noise measurement, however, enables us to discriminate between $e$ and $e/3$. These beautiful experiments in 1997 gave the direct consequence of the existence of the fractional charge. Note that the Nobel Prize in physics was awarded for the fractional quantum Hall effect in 1998 just after these reports. It is also worth mentioning that the shot noise due to $e^* = 2e$, namely the Cooper pair, was observed experimentally in the super-normal junction~\cite{JehlNature2000}.

More recently, the shot noise serves as an important tool to probe the quantum many-body effects such as the Kondo effect. Recent theory predicts an interesting phenomenon; the electron transport through the quantum dot in the Kondo regime exhibits the electron bunching, that is, two electrons are simultaneously scattered because of the Kondo correlation at the quantum dot~\cite{SelaPRL2006,GolubPRB2006,GogolinPRL2006,MoraPRL2008,%
MoraPRB2009_2,SakanoPRB2011}. This prediction was experimentally addressed by several groups including us~\cite{ZarchinPRB2007,DelattreNatPhys2009,YamauchiPRL2011} and very recently we finally confirmed the theoretical prediction quantitatively~\cite{FerrierNatPhys2016}. In this recent work, through a precise shot noise measurement, we show that the Landauer-B{\"u}ttiker formalism is not sufficient to describe the many-body effect in the nonequilibrium regime. Such an experimental attempt may be viewed as a ``collision experiment'' on a chip, where the ``internal structure'' of the exotic Kondo state is detected by observing how electrons are scattered by that state. 

Entanglement (nonlocal quantum correlation) is a key concept in quantum information technology research field. There are several theoretical proposal to create and detect the entanglement in solid state devices by means of the shot noise measurement (for example, refer Refs.~\cite{KawabataJPSJ2001,BeenakkerPRL2003,SamuelssonPRL2004}). The central idea is to detect the violation of Bell's inequality through the correlation between the current fluctuations at two spatially separated points. Towards this direction, two-particle interference  was demonstrated in  the Hanbury-Brown-Twiss interferometer~\cite{NederNature2007}. 
Several recent experiments along this direction with single electron emission~\cite{BocquillonScience2013,DuboisNature2013,FreulonNatComm2015} also nicely rely on the shot noise measurement.
While these attempts have strong analogy with those extensively performed in quantum optics field, such effort to create and detect the entanglement in electronic systems would bring us unique opportunities as the electronic systems have variety of many-body effects that do not exist in photonic systems.

Many studies on the noise in mesoscopic systems were reported especially these several years and therefore the references introduced above constitute only a partial list. ``The noise is the signal'' is the title of the article written by Landauer in 1998~\cite{LandauerNature1998}. Although this sounds contradictory, the studies introduced above surely prove that essential information can be obtained through the noise measurement. 

\subsection{How to Measure Noise}

Here we explain some technical aspects of the noise measurement. The noise measurement in mesoscopic systems is technically more difficult than the conventional conductance measurement. We started the development of the noise measurement technique in 2005 and continue working on it until now~\cite{HashisakaJPC2008,HashisakaPSS2008,HashisakaRSI2009,ArakawaAPL2013}. 

\begin{figure}[tbp]
\center\includegraphics[width=.99\linewidth]{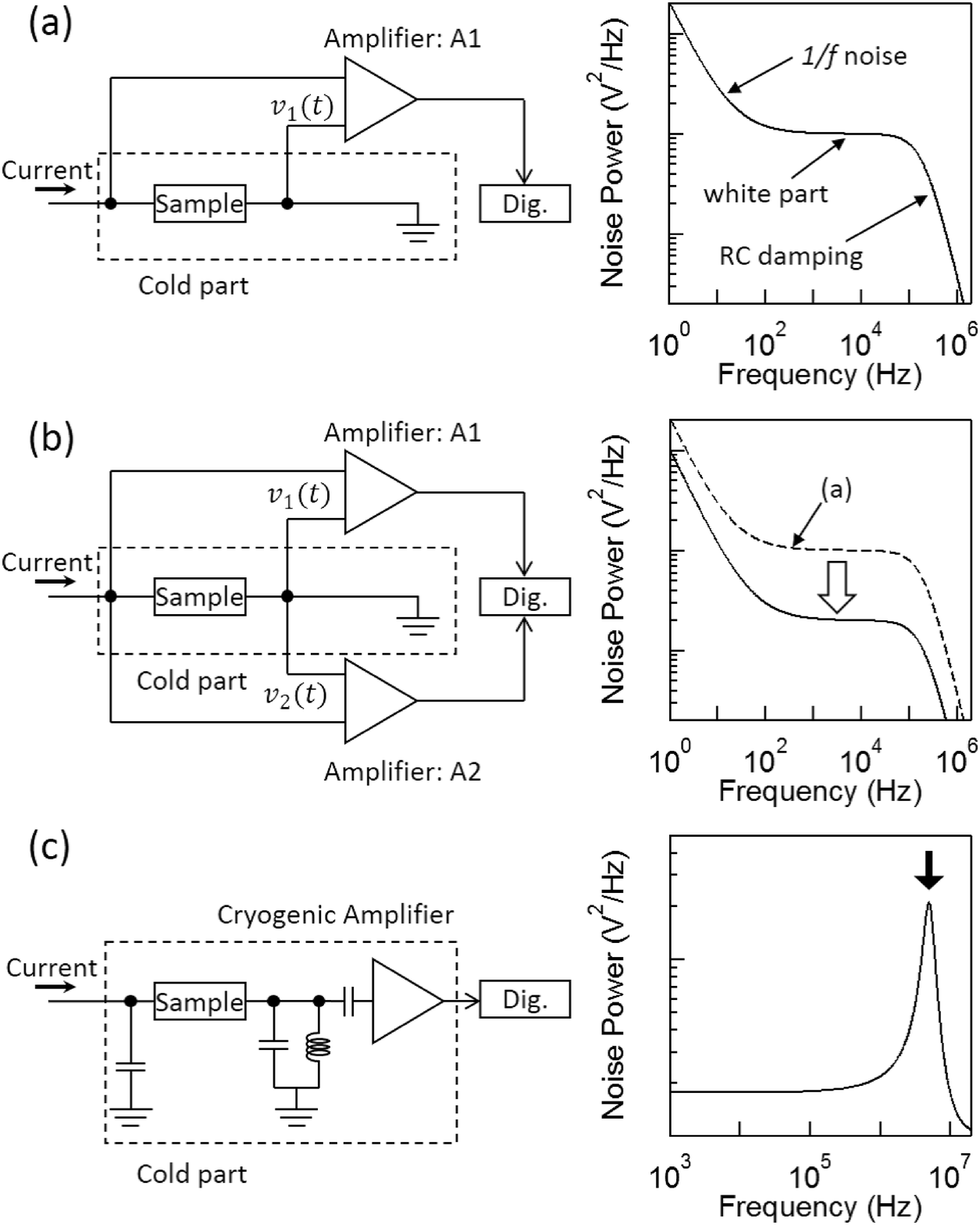}
\caption{(a) Straightforward setup for the voltage noise measurement of the sample placed in the cold part (typically less than 1~K) (left) and the expected voltage noise power spectral density (right). The voltage signal $v_1(t)$ is recorded at the digitizer (Dig.) through the amplifier (A1). (b)  Setup for the voltage noise measurement by using the cross-correlation method (left) and the expected voltage noise power spectral density, where the spectrum (a) is superposed with a dashed curve. By virtue of the cross-correlation, the background due to the extrinsic noise is reduced as shown by the arrow.  (c) Setup for the voltage noise measurement by using $LC$ circuit to increase the measurement band width. The  noise only at a specific frequency can be measured as indicated by the arrow.  }
\label{noiseexp}
\end{figure}

A measurement principle is schematically shown in Fig.~\ref{noiseexp} (a). Usually, the mesoscopic devices should be cooled down below 1~K in order to observe coherent transport. Instead of measuring the current noise spectral density $S$, we measure the voltage noise spectral density $S_V$ since voltage amplifiers are more common and easier to handle than current amplifiers. The time-dependent voltage signal across the sample $v_1(t)$ is recorded at the digitizer (Dig.) through the amplifier (A1)  and is converted to the voltage spectral density $S_V$  via the fast Fourier transformation (FFT). A typical  spectrum is shown in the right panel of Fig.~\ref{noiseexp} (a). In the low-frequency limit, $S_V \propto  \langle (\delta v_1(t))^2 \rangle$, where $\delta v_1(t) \equiv v_1(t) - \langle v_1(t) \rangle$. By using the conductance $G$ of the sample, we get $S = G^2S_V$. 

Thus, to measure the current noise seems, in principle, straightforward. However, there are several technical difficulties. As already shown in Fig.~\ref{noiseexp} (a), the measured spectrum usually consists of three contributions. First, the $1/f$ contribution is dominant at low frequency, which comes from the amplifier and the device itself. The contribution of such $1/f$ noises is very often not negligible below a few kHz. At higher frequencies, we see the frequency-independent flat contribution, which consists of two components. The one is the intrinsic white noise from the device, which we want to measure, and the other is the external noise such as the amplifier noise and the noise of the surrounding environment coming through the measurement cables. Sometimes, these external noises appear as spurious peaks in the spectrum. At much higher frequencies, the signal is damped due to the $RC$-damping effect, which is inevitably caused by the device resistance and the capacitance between the device and the other parts, say, the ground.   The resistance of typical mesoscopic conductors is of the order of the inverse of the conductance quantum, namely $\sim 10$~k$\Omega$ and the measurement band width is reduced below several ten kHz, because the measurement cables, whose lengths are determined by the size of the refrigerator, have the total capacitance of several hundred pF. In this way, the frequency-independent component is often difficult to obtain when the $1/f$ noise is large, and also, even if we observe the frequency-independent contribution, it is often deteriorated by the external noises. 

To overcome the above difficulties, we implement the cross-correlation noise measurement~\cite{KumarPRL1996,SampietroRSI1999}. As shown in Fig.~\ref{noiseexp} (b), we use two amplifiers (A1 and A2). The voltage signals simultaneously detected at A1 and A2 are expressed as $v_1(t) = v_{int}(t) + v_{1, ext}(t)$ and $v_2(t) = v_{int}(t) + v_{2, ext}(t)$, respectively. Here $v_{int}(t)$ is the intrinsic signal from the device while $v_{i, ext}(t)~(i=1, 2)$ is the extrinsic signal coming from the amplifiers and the measurement lines for A1 and A2, respectively. At the digitizer, we calculate the cross-correlated spectral density between the two signals $v_1(t)$ and $v_2(t)$. By taking the long-time averaging for this spectral density, we expect that $S_V \propto \langle \delta v_1(t) \delta v_2(t) \rangle \rightarrow \langle (\delta v_{int}(t))^2 \rangle$ as the extrinsic signals $v_{1, ext}(t)$ and $v_{2, ext}(t)$ are not correlated each other. As a result, we obtain the spectrum as shown in the right panel of Fig.~\ref{noiseexp} (b). The spectrum is usually more suitable for the noise analysis than the spectrum shown in Fig.~\ref{noiseexp} (a). 

The advantage of this cross-correlation method is that we can use the commercial voltage amplifiers at room temperature, which have the well-defined stable performance if we can choose reliable ones. Thus this method is often used in mesoscopic field as seen in, for example, Refs.~\cite{KumarPRL1996,SaminadayarPRL1997,ArakawaPRL2015}. However, as this method requires the long-time averaging to reduce the extrinsic noises, the measurement is time-consuming. Also, if the $1/f$ noise from the sample is non-negligible, the measurement itself is eventually impossible.

To be essentially free from the $1/f$ noise, it is necessary to measure the noise at much higher frequency. To this end, we use the inductor-capacitor ($LC$) resonant circuit by using  a home-made cryogenic amplifier as shown in %
Fig.~\ref{noiseexp}~(c)~\cite{de-PicciottoNature1997,DiCarloRSI2006,HashisakaJPC2008,HashisakaPSS2008,ArakawaAPL2013}. First, the current noise occurring in the device is converted to the voltage noise at the $LC$ circuit, whose resonant frequency is set around a few MHz in our case. The purpose of the $LC$ circuit is to increase the measurement band width. To precisely obtain the current noise, we often need a long-time averaging, which means that the measurement requires a lot of time when it is performed at frequency as low as a few kHz. On the other hand, by using the $LC$ circuit we can increase the band width from a few kHz to a few MHz, which indicates that the measurement time can be one thousandth shorter. At the same time, by increasing the band width in this way, we expect to reduce the influence of the $1/f$ noise, say, by one thousandth. The disadvantage of using this technique, on the other hand, is that the noise only at a specific frequency can be measured and the full spectra cannot be obtained. Therefore, we have to take great care to investigate whether or not the obtained signal is frequency-dependent component by cautiously looking at the temperature- and bias- dependence of the measured signal. 

In order to prevent the rise of the electron temperature due to the room temperature environment through the measurement lines, high frequency filters and the thermal anchors should be used. We are now able to perform the noise measurement with the electron temperature as low as 20~mK.  Technical details are given in Refs.~\cite{DiCarloRSI2006,HashisakaJPC2008,HashisakaPSS2008,HashisakaRSI2009,ArakawaAPL2013}.

%==========================================================================

\section{Experimental Test of Fluctuation Theorem in Quantum Regime}
\label{FTtest}

\subsection{Fluctuation Theorem in Electron Transport}

We have discussed the significance of the fluctuation (or noise) in statistical physics in Sec.~\ref{IntroNoise} and the interest of the noise measurement in mesoscopic physics in Sec.~\ref{IntroMesoNoise}.  Now we combine these two and see how the noise measurement in mesoscopic systems contributes to nonequilibrium statistical physics. In this Section, we discuss the experimental test of the fluctuation theorem (FT). 
Actually, FT is expected to play an important role in understanding transport in mesoscopic systems~\cite{AndrieuxJCP2004,TobiskaPRB2005,EspositoPRB2007,SaitoPRB2008,ForsterPRL2008,%
AndrieuxNJP2009,EspositoRMP2009,UtsumiPRB2009,UtsumiPRB2010,CampisiPRL2010,AltlandPRB2010,CampisiRMP2011}.
First, let us look how FT is relevant in quantum transport~\cite{SaitoPRB2008,UtsumiPRB2009,NakamuraPRB2011}.

\begin{figure}[tbp]
\center \includegraphics[width=.99\linewidth]{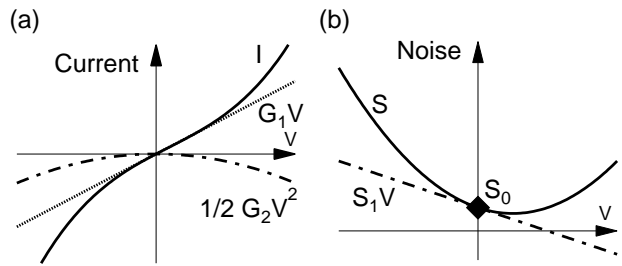} 
\caption{ (a) Current-voltage characteristic as a
function of the bias voltage $V$. Ohm's law holds around $V = 0$,
while the current  is nonlinear at high bias regime and the $I-V$ characteristics
has a polynomial form of $V$ with finite coefficients 
$G_2$, $G_3, \dots$ as in Eq.~(\ref{CurrentPoly}).   
 This graph presents the total current $I$, the $G_1V$
contribution, and the $1/2! G_2 V^2$ contribution, in the solid, dashed,
and dot-dashed curves, respectively.   (b) The current noise  $S$ (shown in
the solid curve) is expressed in a polynomial form of $V$ as in
Eq.~(\ref{NoisePoly}). The FD relation tells that $S(V = 0) =
S_0 = 4k_BTG_1$.}
\label{Coeff}
\end{figure}

Consider a two-terminal device. We also assume that there is no magnetic field. As shown in Fig.~\ref{Coeff}(a), when the bias voltage $V$ is applied to the device, the current $I$ is generated such that it can be expressed as,
\begin{equation}
I = G_1 V + \frac{1}{2!}G_2 V^2 + \frac{1}{3!}G_3 V^3 + \cdots .
\label{CurrentPoly}
\end{equation}
Here, $G_1$ is the conductance, which is exactly the same as $G$ as we introduced in Eq.~(\ref{JN}). If $I = G_1 V $ is the case, the system is in the linear regime and the conventional linear response theory is applicable. However, when $\vert eV \vert > k_BT$ holds, the system is in the nonequilibrium regime and the linear response theory is not applicable. In such a situation, the higher order coefficients $G_2$, $G_3, \dots$ play a critical role. 

Now consider the current noise $S$. In the same way, as presented in Fig.~\ref{Coeff}(b), $S$ is expressed by 
\begin{equation}
 S = S_0 + S_1 V + \frac{1}{2!}S_2 V^2 + \cdots .
\label{NoisePoly}
\end{equation}
Here $S_0$ is the thermal noise. As we discussed in Eq.~(\ref{JN}), the following FD relation holds;
\begin{equation}
 S_0 = 4 k_B T G_1.
\label{1st}
\end{equation}
This is the relation between the coefficients of the first term of Eq.~(\ref{CurrentPoly}) and Eq.~(\ref{NoisePoly}). 

Now, it might be tempting to ask whether or not such a relation as Eq.~(\ref{1st}) exists for the coefficient of the higher order terms. The answer is yes.  By using  FT,  the following nontrivial relation was predicted~\cite{SaitoPRB2008},
\begin{equation}
S_1= 2 k_B T G_2.
\label{2nd}
\end{equation}
This relation connects between the nonlinear response characterized by $G_2$ and the nonequilibrium fluctuation characterized by $S_1$. Therefore, it is beyond the FD relation that is only valid in the linear response regime around the equilibrium. 

The reason for this relation can be explained as follows~\cite{SaitoPRB2008,NakamuraPRB2011}. FT is the exact relation to connect between the possibility of the entropy increase and that of the entropy decrease. A similar relation holds true in the electron transport. The electron transport occurs in a ``lead-device-lead'' system and it can be viewed as an electron exchange process between the two reservoirs via the device. When there is a voltage difference $V$ between the two reservoirs, the difference in the chemical potential is $eV$. Electrons go back and forth between the two reservoirs. Now, consider the probability $P(Q)$ that $Q$ electrons move from the left lead to the right one. By using the time-reversal symmetry, the particle number conservation, and the energy conservation, one can show the following relation~\cite{NakamuraPRB2011,SaitoPRB2008,UtsumiPRB2009},
\begin{equation}
\frac{P(Q)}{P(-Q)} =  \exp\left(\frac{eV}{k_BT}Q\right).
\label{QT_FT}
\end{equation}
This is FT in the electron transport. 

The relevance between the above FT and the original FT described by Eq.~(\ref{FTeqn}) is as follows. The original FT connects the probability of a process and that of the time-reversed counterpart in terms of the entropy production. In electron transport, the entropy production is related to Joule heating. When $Q$ electrons transmit from the one reservoir to the other across the potential difference of $eV$, the Joule heating of $QeV$ occurs after they equilibrate with the thermal bath at temperature $T$. The corresponding entropy production is $QeV/T$, which appears in the right hand side of Eq.~(\ref{QT_FT}). This connects the probability of the $Q$-electron transfer $P(Q)$ and that of ``$-Q$''-electron transfer $P(-Q)$ in a similar manner as in Eq.~(\ref{FTeqn}). 

Eq.~(\ref{QT_FT}) poses a very strong constraint in the electron exchange process. When there is no potential difference between the two leads, $\langle Q \rangle =0$ is expected naturally. From Eq.~(\ref{QT_FT}), $P(Q) = P(-Q)$ holds, telling us that the possibility that $Q$ electrons move from the left to the right and the possibility of the opposite case are the same. This is a matter of course in the equilibrium state. Interestingly, the FD relation, namely, Eq.~(\ref{JN}) or Eq.~(\ref{1st}) is easily deduced from this fact. The current $I$ and the current noise $S$ are the expectation of the number of electrons and the expectation of the variance of the number of electrons that are exchanged for a finite time $\tau$, respectively. Therefore, the followings hold.
\begin{equation}
\frac{I}{e}=\frac{\langle Q \rangle}{\tau},
\label{Expect_Q}
\end{equation}
and
\begin{equation}
\frac{S}{e^2} = 2\frac{\langle (Q-\langle Q \rangle)^2\rangle}{\tau} = 2\frac{\langle Q^2
\rangle - \langle Q\rangle^2}{\tau}.
\label{NoiseVariance}
\end{equation}
Now, $A \equiv eV/k_BT$ is defined, 
\begin{eqnarray}
\langle Q \rangle &=& \sum_Q Q P(Q) =
- \sum_Q Q P(Q) e^{-A Q} \nonumber \\
&=&
-\langle Q \rangle + A \langle Q^2 \rangle -{A^2\over 2! }\langle Q^3 \rangle +
\cdots. 
\label{FTTaylor}
\end{eqnarray}
is deduced. Here FT (Eq.~(\ref{QT_FT})) was used. $e^{-AQ}$ was expanded in a Taylor series with regard to $AQ$. We also expand $\langle Q^n \rangle ~(n= 1, 2 \cdots)$ in a Taylor series as follows. 
\begin{equation}
\langle Q^n \rangle = \langle Q^n \rangle_0 + A \langle Q^n \rangle_1
+ {A^2\over 2!} \langle Q^n \rangle_2 + \cdots
\label{FTTaylor_n}
\end{equation}
The coefficients of the term $A^k$ are expressed as $\langle Q^n \rangle_k \quad (k=0, 1, 2 \cdots)$. 
For example, 
\begin{eqnarray}
\frac{I}{e} =\frac{\langle Q \rangle}{\tau}  
&=& \frac{1}{\tau} \left(\langle Q \rangle_0  +  \langle Q \rangle_1 A   + \cdots \right)  \\
&=& \frac{1}{\tau} \left(\langle Q \rangle_1 \frac{eV}{k_BT}   + \cdots \right),
\end{eqnarray}
since it is easily proved that $ \langle Q \rangle_0=0$ by using Eqs.~(\ref{FTTaylor}) and (\ref{FTTaylor_n}). 
Note that this relation corresponds to the polynomial $I = G_1V + \cdots$ with $G_1 = (\langle Q
\rangle_1 / k_BT) (e^2/\tau)$.
A similar analysis is performed for Eq.~(\ref{NoiseVariance}) and $S_0 = 2 \langle Q^2 \rangle_0 (e^2/\tau)$ is obtained since $ \langle Q \rangle_0=0$.
By comparing the coefficients of the first order term in $A$, 
\begin{equation}
\langle Q^2 \rangle_0 = 2 \langle Q \rangle_1.
\end{equation}
This is exactly the FD relation shown in Eq.~(\ref{1st}), or equivalently Eq.~(\ref{JN}). The left hand side is the first order of the fluctuation while the right hand side is the coefficient to describe the linear response. 

In the same way, by considering the coefficients of $A^2$, we obtain $\langle Q^2 \rangle_1 = \langle Q \rangle_2$. This leads to Eq.~(\ref{2nd}). Theoretically, not only between $G_2$ and $S_1$, but also among the other coefficients of the higher order terms ($G_3, G_4 \cdots$ and $S_2, S_3 \cdots$), numerous relations are deduced systematically~\cite{SaitoPRB2008}. However, for the higher order cases, one has to treat the higher order fluctuations such as $\langle (\delta I)^3 \rangle$, $\langle (\delta I)^4 \rangle, \dots$. Experimentally, to measure such higher order fluctuations (namely, higher order cumulants) beyond the second order one is difficult and therefore in this Review we only focus on the relations Eqs.~(\ref{1st}) and (\ref{2nd}).

As FT is based on the time-reversal symmetry of the system, the magnetic field, which affects this symmetry,  should be explicitly treated. Therefore,  in the presence of the magnetic field ($B$), the probability $P(Q,B)$ should be considered. In this case, FT is shown to have the following form~\cite{SaitoPRB2008,UtsumiPRB2009},
\begin{equation}
\frac{P(Q,B)}{P(-Q, -B)} = \exp(AQ).
\label{PQB}
\end{equation}
Therefore, we now consider the symmetric and antisymmetric parts of $P(Q,B)$ in a way such that $P_{\pm} (Q) \equiv P(Q,B) \pm P(Q,-B)$. By definition, 
\begin{equation}
P_\pm(Q) = \pm P_{\pm}(-Q)e^{AQ}
\end{equation}
Accordingly, we consider these two coefficients,
\begin{equation}
G_2^{S, A} (B)\equiv G_2 (B) \pm G_2 (-B)
\label{G2_SA}
\end{equation}
and
\begin{equation}
S_1^{S, A} (B)\equiv S_1 (B) \pm S_1 (-B).
\label{S1_SA}
\end{equation}
Note that take $+$ for $S$ (symmetric) and $-$ for $A$ (antisymmetric).
$P_+$ is nothing but $P(Q)$ in the zero-field case and therefore
\begin{equation}
S_1^S = 2k_BTG_2^S
\label{FT_Sym}
\end{equation}
holds. However, for $P_-$, 
\begin{equation}
S_1^A = 6k_BTG_2^A
\label{FT_Asym}
\end{equation}
can be deduced in the same way as above. It is significant that the symmetric
and antisymmetric components have different numerical factors. 

We make a comment on the Onsager-Casimir reciprocal relations, which is the fundamental property of the systems around the equilibrium.  
In two terminal conductors, the reciprocity results in
\begin{equation}
G_1(B) = G_1(-B).
\end{equation}
Interestingly, this relation can be derived by using Eq.~(\ref{PQB})~\cite{SaitoPRB2008,UtsumiPRB2009}.
This means that Eq.~(\ref{PQB}) already includes this important reciprocity. Of course, in the nonequilibrium regime, such reciprocity is not necessarily valid. Actually, generally, $G_k(B) \neq G_k(-B)~(k\geq2)$ and we will see that $G_2^{A} \neq 0$ in our experiment below.

\subsection{Experiment}

Now let us explain our experiment to show the validity of the above expected relations (Eqs.~(\ref{2nd}), (\ref{FT_Sym}), and (\ref{FT_Asym})) in the quantum regime. To this end, we used an electronic interferometer, or Aharonov-Bohm (AB) ring~\cite{LeturcqPRL2006,YamauchiPRB2009}, as a typical mesoscopic conductor. While other mesoscopic systems, such as a quantum dot or a quantum wire, can be used for the same purpose, an AB ring has an advantage in that it is easy to prove that the coherent transport takes place as signaled by the AB oscillation. 

\begin{figure}[tbp]
\center \includegraphics[width=.9\linewidth]{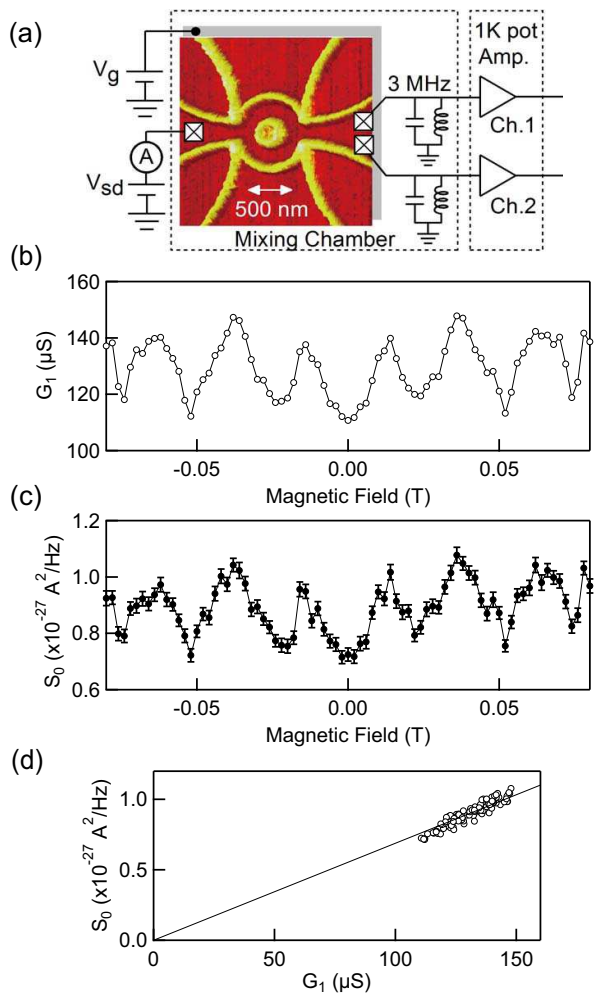} 
\caption{(a) Atomic force microscope (AFM) image of the AB ring fabricated by local oxidation using an AFM~\cite{HeldAPL1998} on a GaAs/AlGaAs 2DEG. Schematic experimental setup for the two-terminal current and noise measurements is also given.  (b) Conductance $G_1$ of the ring as a function of $B$ obtained at $V_{\rm g} = -0.09$~V. (c) Corresponding equilibrium noise $S_0$ as a function of $B$. $G_1$ and $S_0$ shown in (b) and (c) were simultaneously measured.  (d) $S_0$ is plotted as a function of $G_1$. This graph is obtained from (b) and (c) by using the magnetic field as an internal parameter.  The solid line indicates the FD relation of  $S_0 = 4 k_BTG_1$ with $T= 125$~mK (see Eq.~(\ref{1st})). }
\label{ABring_EqCond}
\end{figure}

The AB ring was fabricated by the AFM (atomic force microscopy) lithography~\cite{HeldAPL1998}. The AFM lithography enables us to make nano patterning on the semiconductor surface by the voltage-biased AFM cantilever  through the local oxidization. When this technique is applied to  GaAs/AlGaAs heterostructure 2DEG, the electrons are depleted beneath the locally oxidized surface. As a result, precise nanofabrication is realized. We fabricated an AB ring with the diameter of 460~nm on the 2DEG, which has the electron density $3.7 \times 10^{11}$~cm$^{-2}$ and the mobility $2.7 \times 10^5$~cm$^2/$Vs with the electron mean free path of 2.7 $\mu$m. Figure~\ref{ABring_EqCond}(a) shows the AFM image of the ring. The 2DEG has a back gate to tune the electron density. The conductance of the AB ring can be modulated by the back gate voltage $V_g$ and by the magnetic field $B$ perpendicularly applied to the 2DEG through the AB effect. 

Figure~\ref{ABring_EqCond}(b) shows the conductance, namely $G_1$ of this conductor, measured at $V_{\rm g} = -0.09$~V. As a function of $B$, the conductance shows a clear AB oscillation with an oscillation period of 25 mT. This value agrees the ring radius of $r = 230$~nm, as $\frac{h}{e}\frac{1}{\pi r^2}=24.6$~mT. The conductance of the ring ranges between 1.3 and 1.7 in units of $2e^2/h$ (12.9 k$\Omega$)$^{-1}$ with typical visibility of $\sim 0.13$. The presence of electron interferences guarantees the coherent electron transport in the device.

In the present experiment, we measure the current-voltage characteristic as well as the noise at each bias voltage. The noise measurement was performed by combining the cross-correlation technique and the  $LC$ resonant circuit method as described in the previous section. We numerically fit the result of the current and the noise by using Eqs.~(\ref{CurrentPoly}) and (\ref{NoisePoly}). The window for the fitting was set such that $\vert eV \vert \lesssim 3 k_B T$, because we use FT for the expansion around $A = eV/k_BT=0$. 
In this bias regime, the noise consists of the thermal noise and the shot noise. The shot noise is the result of the stochastic processes of the transmission and reflection at the AB ring as discussed in Sec.~\ref{subsec:exampleofnoise}. Generally, the transmission probability depends on the electron energy and may be affected by the electron-electron interaction, which is also the case in the present AB ring as we see below. While such a  situation is beyond the conventional shot noise formula Eq.~(\ref{ShotTheory})  (see the assumption for Eq.~(\ref{ShotTheory})), the present treatment based on FT is still valid. On the other hand, we treat the coefficients in the Taylor expansion, which means that we cannot treat a far-from-equilibrium situation but addresses the crossover from the equilibrium to the nonequilibrium around $A = eV/k_BT=0$.

\subsection{Results and Discussion}

We first present the equilibrium noise. Corresponding to the conductance plot shown in Fig.~\ref{ABring_EqCond}(b), the noise  obtained at $V=0$ (namely, $S_0$) is plotted as a function of $B$ in Fig.~\ref{ABring_EqCond}(c). The behavior of $S_0$ follows that of $G_1$ as expected from the FD relation (Eq.~(\ref{1st})). The proportionality between $G_1$ and $S_0$ shown in Fig.~\ref{ABring_EqCond}(d) tells that $S_0 = 4 k_B T G_1$ holds with the electron temperature $T=125$~mK. In addition to this temperature, we performed the measurement at 300~mK and 450~mK and confirmed the FD relation at each temperature.

The remarkable observation that should be emphasized in  Figs.~\ref{ABring_EqCond}(b) and (c) is that both the conductance $G_1$ and the thermal noise $S_0$ are nicely symmetric with regard to the magnetic field reversal such that $G_1(B) = G_1(-B)$ and  $S_0(B) = S_0(-B)$. This is the manifestation of the Onsager-Casimir relation, which characterizes the response of a system around the equilibrium. 

\begin{figure}[tbp]
\center \includegraphics[width=0.7\linewidth]{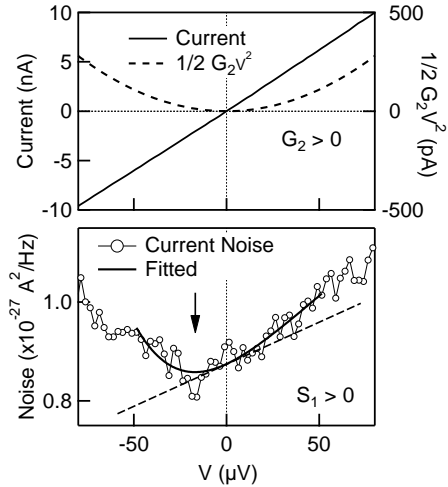} 
\caption{(Upper panel) Example of $I$-$V$ characteristics, where $\frac{1}{2}G_2V^2$ component deduced from the fitting is superposed as a dashed curve. The data was obtained at the magnetic field $B = 0$~T, $T = 125$~mK, and $V_g = 0.04$~V. (Lower panel) Corresponding $S$, where the results of the polynomial fitting and $S_1V$ component are superposed by solid and dashed curves, respectively. The arrow shows that the noise takes its minimum at a finite bias voltage.}
\label{NegativeExcNoise}
\end{figure}

When the voltage is applied to the AB ring, the nonequilibrium noise emerges. An example is shown in Fig.~\ref{NegativeExcNoise}, where the upper and lower panels show the current and the current noise as a function of $V$, respectively. In the upper panel, the raw data of the current and the component of $1/2 G_2 V^2$ that is obtained in the numerical fitting are plotted in the left and the right axes, respectively. There exists small but finite contribution of $G_2$, reflecting the nonequilibrium nonlinear nature of the conduction. The lower panel shows the obtained noise (open circles) with the result of the numerical fitting (solid curves). Clearly, the noise takes its minimum at finite bias voltage instead of $V=0$. While this would be expected when $S_1$ is finite, the result gives a nice example of negative excess noise, that is the nonequilibrium noise smaller than the equilibrium one~\cite{LesovikZPB1993,BlanterPR2000}. 

\begin{figure}[tbp]
\center \includegraphics[width=.99\linewidth]{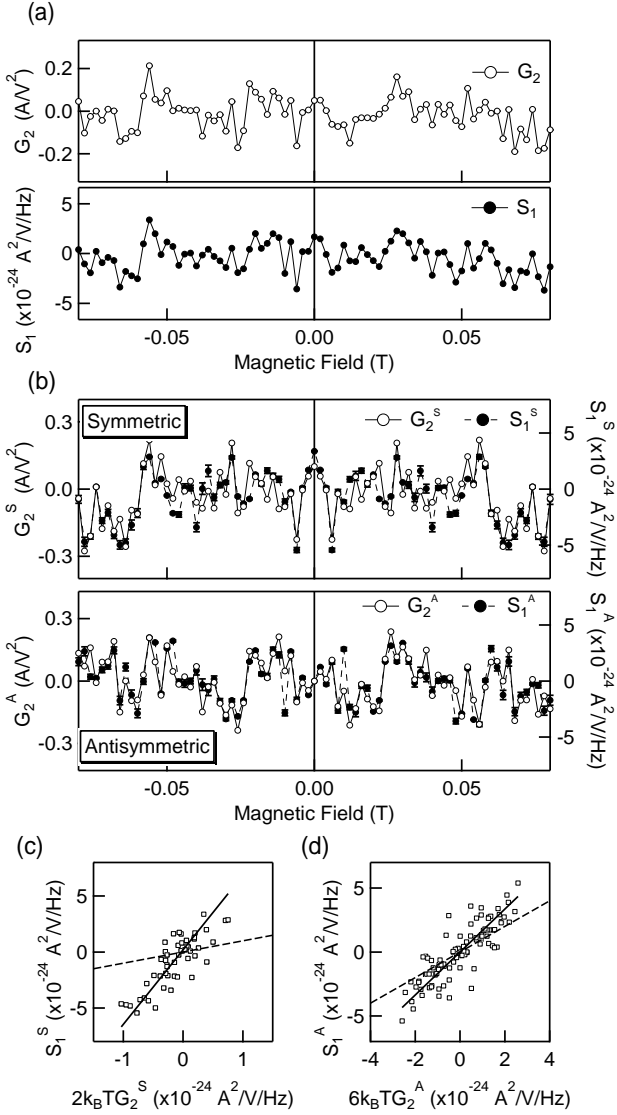} 
%\vspace{3cm}
\caption{(a) Second-order coefficients $G_2$ (upper panel) and $S_1$ (lower panel) as a function of $B$. 
(b) (upper panel) Symmetric components $G_2^S$ and $S_1^S$ as a function of $B$.  (lower panel) Anti-symmetric components  $G_2^A$  and $S_1^A$ as a function of $B$.  (c) $S_1^S$ as a function of $2k_BTG_2^S$.  (d) $S_1^A$ as a function of $6k_BTG_2^A$. In (c) and (d), the solid line and the dashed line are the result of the fitting and the theoretical prediction, respectively.}
\label{S1G2result}
\end{figure}

Now, we investigate the behavior of $G_2$ and $S_1$ as a function of $B$, which is shown in the upper and lower panels of Fig.~\ref{S1G2result}(a), respectively. Note that these data were obtained simultaneously with the data shown in Figs.~\ref{ABring_EqCond}(b) and (c). As we mentioned already,  $G_1$ and $S_0$ are symmetric with regard to the magnetic field reversal. In contrast, $G_2$ and $S_1$ has no such symmetry. In other words, the Onsager-Casimir relation does not hold. It may sound natural, as $G_2$ and $S_1$ characterize the system in the nonlinear nonequilibrium regime. However, it is experimentally significant to see this clear transition from the equilibrium regime to the nonequilibrium one, because such a well-defined system allows us to precisely connect between the two regimes.

What is the microscopic mechanism to break the Onsager-Casimir reciprocity in the nonlinear transport regime? Actually this phenomenon was known before~\cite{SanchezPRL2004,SpivakPRL2004,WeiPRL2005,LeturcqPRL2006} and is attributed to electron-electron interactions in mesoscopic conductors~\cite{SanchezPRL2004,SpivakPRL2004}. As the magnetic field changes, the electronic property inside the ring and hence $G_2$ and $S_1$ are modulated at finite bias through the interaction. In the Landauer picture for non-interacting electrons~\cite{BlanterPR2000}, the transmission probability of electrons, which is defined at the equilibrium, is to obey the Onsager-Casimir reciprocity regardless of applied bias voltages. Therefore the breaking of this reciprocity directly indicates that the system behaves in a nontrivial way in the nonequilibrium regime and that the transport can no more be explained in the single-particle picture. 

To be more quantitative, in Fig.~\ref{S1G2result}(b), the symmetric component ($G_2^S (B)$ and $S_1^S (B)$) and the antisymmetric component ($G_2^A (B)$ and $S_1^A (B)$) of $G_2$ and $S_1$ are shown in the upper and lower panels, respectively (see Eqs.~(\ref{G2_SA}) and (\ref{S1_SA})).  As discussed above, while the Landauer picture simply predicts $G_2^A (B) =0$, we observe a finite $G_2^A (B)$ as large as $G_2^S (B)$, signaling the significant interaction. Moreover, as seen in the upper panel of Fig.~\ref{S1G2result}(b) the behavior of $G_2^S$  and that of $S_1^S$  follow each other. This is also the case for the antisymmetric component shown in the lower panel of Fig.~\ref{S1G2result}(b). This experimental observation that the nonlinear response strongly correlates with the nonequilibrium fluctuation supports the validity of the relations predicted by FT. 

FT predicts that $S_1^S = 2k_BTG_2^S$ and $S_1^A = 6k_BTG_2^A$ (Eqs.~(\ref{FT_Sym}) and (\ref{FT_Asym})). Figures~\ref{S1G2result}(c) and (d) are the plots of $S_1^S$ vs. $2k_BTG_2^S$ and $S_1^A$ vs. $6k_BTG_2^A$, where the magnetic field is an internal parameter, respectively. Although the data are slightly scattered, we can see the remarkable linearity between the two in both graphs.  Here, the dotted lines are the predicted ones.  The symmetric part (Fig.~\ref{S1G2result}(c)) deviates from the theory while the anti-symmetric part (Fig.~\ref{S1G2result}(c)) is in a better agreement with the theory. 

To make a quantitative comparison between the experimental result and the theory, the Passing-Bablok (PB) regression, rather than the conventional least-square linear fitting, is suitable.  The reason is because both $S_1^S$ and $G_2^S$ (or both $S_1^A$ and $G_2^A$) have statistical uncertainties. In such a situation, the PB regression can statistically estimate the error bars of the coefficients between the two values better than the conventional linear fitting.  For the presented data set, the PB regression~\cite{PassingJCCCB1983} yields
\begin{equation}
S_1^S =6.15^{+1.01}_{-0.78} \times 2k_BTG_2^S
\end{equation}
and
\begin{equation}
S_1^A =
1.68^{+0.22}_{-0.16} \times  6k_BTG_2^A,
\end{equation}
where the error bars give the 95 \% confidence interval. The reason for the observed deviation from the theory is not yet clear, unfortunately. Nevertheless, it is important that the higher-order correlation, which is theoretically predicted, exists with different coefficients ($S_1^A/S_1^S \sim 0.82 \times G_2^A/G_2^S$). It is important to add that the values of $S_1^S/G_2^S$ and $S_1^A/G_2^A$ were found to be linearly dependent on the temperature~\cite{NakamuraPRL2010} as was theoretically expected.

Finally, we mentioned  the direct test of the validity of the microreversibility. The necessary condition for FT  to hold is the microreversibility or the detailed balance. Several years ago, it was theoretically suggested that the microreversibility might not hold in the presence of the magnetic field~\cite{ForsterPRL2008}. Interestingly, our experimental data on the anti-symmetric component offers an opportunity to test this possibility and we proved that the microreversibility is maintained within the experimental precision~\cite{NakamuraPRB2011}. It is interesting to see that one can directly address a fundamental problem in statistical physics by means of the noise measurement. 

\section{Conclusion and Perspective}
\label{SecConcPersp}
In this Review, we overview the research on fluctuation (or noise) and discuss our experiment on FT in the quantum regime. Our experiment was the first attempt to tackle this issue. At this moment, we have not yet fully proven the theorem, but we could show that FT is semi-quantitatively valid in the description of quantum transport in mesoscopic systems, which gives the relation beyond the conventional linear response theory. The significance of our work lies in the observation that there surely exists a relation between the nonlinear response and the nonequilibrium fluctuation. 

In addition, the breaking of the Onsager-Casimir reciprocity, namely the nonlinear conductance and fluctuation that are not symmetric with regard to the magnetic field reversal, is definitely nontrivial beyond the Landauer picture. This fact indicates that FT conveys us essentially new information on quantum transport.  Of course, this fact does not degrade the Landauer picture, which has played and is still playing a central role in mesoscopic transport. The Landauer picture allows us to treat the transport as quantum scattering problems, where the conductance is linked with the transmission probability. On the other hand, the description based on FT does not give us such a microscopic picture but the variation of the number of the electrons in the reservoirs is correctly taken into account (remember that, in the Landauer picture, the reservoir can emit and absorb infinite number of electrons). Therefore, both descriptions are complementary to each other. We believe that by combining these two pictures, nonequilibrium properties in mesoscopic systems in the presence of interaction effects will be further addressed.

By pursuing this research direction, we can experimentally address several fundamental problems associated with both quantum mechanics and statistical physics, and will shed new light in  nonequilibrium statistical physics. One could call such research field mesoscopic nonequilibrium statistical physics. For example, due to the observation problem peculiar to quantum physics, the foundation of the linear response theory has been discussed in 1950's~\cite{TakahashiJPSJ1952}. Now, however, precise noise measurements may allow us to experimentally answer such old important problems. In the classical case, several experiments closely related to Maxwell's demon were demonstrated as FT has been expanded to include information~\cite{ToyabeNatPhys2010,KoskiPRL2015,ParrondoNatPhys2015}. Is it possible to make a quantum Maxwell's demon? We expect that in near future we can ``observe'' the motion of a single electron with quantum coherence maintained and the connection between quantum fluctuation and FT can be explored. Such an attempt will serve as bridge between quantum physics, statistical physics, and thermodynamics.

\section*{Acknowledgements}
We greatly thank many collaborators involved in the experimental works described in this Review. Especially, for the experiment on FT, we appreciate the fruitful collaboration with Shuji Nakamura, Yoshiaki Yamauchi, Masayuki Hashisaka, Kensaku Chida, Teruo Ono, Renaud Leturcq, Klaus Ensslin, Keiji Saito, Yasuhiro Utsumi, and Arthur C. Gossard. In addition, Rui Sakano, Akira Oguri, Sadashige Matsuo, Meydi Ferrier, Tomonori Arakawa, Takahiro Tanaka, Yoshitaka Nishihara, Tokuro Hata, and Shota Norimoto are acknowledged for continuous collaboration over several years. 

This work was partially supported by a Grant-in-Aid for Scientific Research (S) (No.~26220711), Grant-in-Aid for Scientific Research on Innovative Areas ``Fluctuation \& Structure'' (No.~25103003) and ``Topological Materials Science'' (KAKENHI Grant No.~15H05854),  and Yazaki Memorial Foundation for Science and Technology. We also acknowledges the stimulating discussions in the meeting of the Cooperative Research Project of RIEC, Tohoku University.

\end{document}